\documentclass[12pt]{article}

\usepackage{color}
\usepackage{ulem}

\usepackage[utf8]{inputenc}
\usepackage{pythontex}
\usepackage{algpseudocode}

\usepackage{amsmath}
\usepackage{amssymb}
\usepackage{amsfonts}
\usepackage{graphicx}
\usepackage{hyperref}
\usepackage{natbib}
\RequirePackage[inline]{enumitem}
\setlist{nosep,font=\bfseries,leftmargin=*,align=left}
\setlist[1]{topsep=\baselineskip,leftmargin=\parindent,labelsep=*,labelwidth=*}
\setlist[enumerate,2]{label={\alph*.},}
\usepackage{appendix}
\usepackage{comment}

\hypersetup{colorlinks,linkcolor={blue},citecolor={blue},urlcolor={red}} 
\usepackage{booktabs}
\usepackage{multirow}
\usepackage{float}
\usepackage{siunitx}
\usepackage{tikz}
\usetikzlibrary{shapes.geometric, arrows, positioning}

\tikzstyle{startstop} = [rectangle, rounded corners, minimum width=3.7cm, minimum height=1cm, text centered, draw=black, text width=3.7cm]
\tikzstyle{startstopnoborder} = [minimum width=3cm, minimum height=1cm, text centered, text width=3cm]
\tikzstyle{decision} = [diamond, minimum width=3cm, minimum height=1cm, text centered, draw=black, text width=3cm]
\tikzstyle{arrow} = [thick,->,>=stealth]

\let\citep\cite
  \let\citet\ocite
\newenvironment{proof}[1][Proof]{\begin{trivlist}
\item[\hskip \labelsep {\bfseries #1}]}{\end{trivlist}}

\newcommand{\runningauthor}[1]{\def\@runningauthor{#1}}
\newcommand{\corraddress}[1]{\def\@corraddress{#1}}
\newcommand{\corremail}[1]{\def\@corremail{#1}}

\newcommand{\fundinginfo}[1]{\def\@fundinginfo{#1}}
\def\@presentaddress{}
\title{\bf An Efficient Image Denoising Method Integrating Multi-resolution Local Clustering and Adaptive Smoothing}
\author{Subhasish Basak, Partha Sarathi Mukherjee\\
    Indian Statistical Institute\\}

\corraddress{Partha Sarathi Mukherjee, Interdisciplinary Statistical Research Unit, Indian Statistical Institute, Kolkata 700108, India}
\corremail{psmukherjee.statistics@gmail.com; psm@isical.ac.in}

\fundinginfo{No funding was received for this research.}

\runningauthor{Basak and Mukherjee}

\begin{document}

  \maketitle

\def\spacingset#1{\renewcommand{\baselinestretch}%
{#1}\small\normalsize} \spacingset{1}


\bigskip
\begin{abstract}
The importance of developing efficient image denoising methods is immense especially for modern applications such as image comparisons, image monitoring, medical image diagnostics, and so forth. Available methods in the vast literature on image denoising can address certain issues in image denoising, but no one single method can solve all such issues. For example, jump regression based methods can preserve linear edges well, but cannot preserve many other fine details of an image. On the other hand, local clustering based methods can preserve fine edge structures, but cannot perform well in presence of heavy noise. The proposed method uses various shapes and sizes of local neighborhood based on local information, and integrates this adaptive approach with the local clustering based smoothing. Theoretical justifications and numerical studies show that the proposed method indeed performs better than these two individual methods, and outperforms many other state-of-the-art techniques as well. This superior performance emphasizes the strong potential of the proposed method for broad applicability in modern image analysis.
\end{abstract}

\noindent%

\spacingset{1.45}
\section{Introduction}
\label{sec:intro}

Easy access to modern image acquisition techniques is making images a popular data format in various disciplines of science. Applications in manufacturing industries include stress and strain analysis of products, anomaly detection in the rolling process, steel and tile surface monitoring, etc. In medical science, various imaging modalities such as X-ray, CT-scan, MRI, and fMRI are being widely used for medical diagnosis. For surveillance of the Earth's surface, satellite images are commonly used. These images serve as a fundamental tool for research in agriculture, forest science, ecology, ecosystems, and many more. Note that image comparisons and sequential image monitoring are important in all of these applications. These images often contain noise, anomalies, etc. that make eventual image analysis unreliable. Therefore, image analysis as a modern-day research area is increasingly gaining a lot of importance. See \cite{Qiu2018QE} for related discussions.

\begin{center}
\begin{figure}[ht!]
\centering
\includegraphics[width=6.5in]{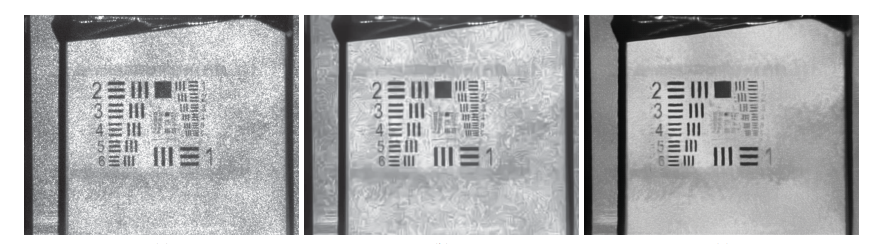}
\caption{Image restoration of real world images provided by \cite{6319405}.}
\label{fig:demo}
\end{figure}
\end{center}

Along with increasing popularity of digital imaging systems, there is a growing demand for near-perfect images. Image denoising is essential in many real-life problems. For example, denoising techniques applied to MRI, CT scan, and X-ray images significantly improve image quality, thereby facilitating more accurate clinical diagnoses. In photography, noise removal enhances the clarity of photographs captured in dark conditions or with high ISO settings. In astronomy, efficient noise removal aids in the identification of celestial bodies and phenomena. Satellite imaging benefits from denoising by refining images captured from space, allowing better analysis and interpretation of geographic and environmental data. Figure \ref{fig:demo} shows the benefits of efficient noise removal from images.
\vspace{-5mm}

An overview of the long history of image denoising can be found in \cite{gonzalezdigital,qiu2005image}. One major type of image denoising algorithms does not detect edges explicitly, but uses the edge information from the observed image intensities. Bilateral filtering methods e.g., \cite{Chu01061998} use the edge information to assign weights in the local intensity averaging procedure. The anisotropic diffusion method proposed by \cite{PeronaMalik1990} uses the edge information to control the direction and the amount of local smoothing. The major drawback of these approaches is that the edges become blurred while smoothing, especially around complicated edge structures such as corners. This is because all pixels in a neighborhood get non-zero weights in the process of local smoothing. Markov random field (MRF) based denoising methods \citep{gemangeman, godtsebas} introduce a line process to use the edge information, whereas methods such as \cite{RUDIN1992259, keeling2003, wangzhou2006}, etc. introduce various penalty terms in appropriate cost minimization problems. These approaches blur the fine image details to some extent. Other popular types of image denoising methods include wavelet transformation methods \citep{portillaetal2003}, non-local means algorithms \citep{buades2005non,buades2010image,liu2008robust}, pointwise shape adaptive methods \citep{foietal2007}, channel or orientation space methods \citep{felsbergetal2006, frankenduits2009} to name a few. One major approach to solve the problem of image denoising is through jump regression analysis \citep{qiu2005image}. The majority of the initial such attempts \citep{QiuSankhya1997, qiu1998discontinuous} detect the edges first, and then use these edge information to denoise the image. Later, \cite{mukherjeeqiu2011} extend similar ideas to denoise 3-D images. 
Jump regression based image denoising techniques that do not explicitly detect edges include methods by \cite{QiuTechnomerics2004, GijbelEtal2006}. See \cite{qiu2007jump} for a detailed discussion. The explicit edge detection based methods do not work well on various types of images such as low resolution images, textured images, etc. To resolve this issue, \cite{mukherjee2015image} propose local pixel clustering based image denoising where two different regions within a local neighborhood are detected by pixel clustering based on image intensity values. An adaptive smoothing method using various shapes and sizes of the neighborhood for local smoothing depending on local edge information is another approach (e.g., \cite{takeda2007kernel}). Even though the neighborhoods are elongated along the edges, they often contain pixels on both sides of edges, resulting in image blurring. In spite of having all these methods, no one method outperforms others on a large class on images. It is to be noted that each of these methods has its specific weaknesses alongside its strengths.

Several types of modern image analyses such as image comparisons \citep{FengQiu2018, roy2024image, GuhaQiuImageComparison}, image monitoring \citep{roy2024control} require feature preserving image denoising as a preliminary step. Sometimes, many such methods use the central ideas of various image smoothing methods as building blocks that suit their own specific requirements. Performance of image comparison and monitoring by such methods depends heavily on the effectiveness of image smoothing. Therefore, it is an urgent requirement that integration of various types of state-of-the-art image denoising methods are done to combine the strengths and remove the weaknesses of the concerned methods as much as possible. Another important aspect to note in this context is that deep learning based denoising methods (e.g., \cite{deepdenoising}) require a large set of images for training or learning. Therefore, such type of denoising methods cannot be easily integrated with image monitoring procedures which itself is a computationally heavy problem. Therefore, requirement of developing integrated denoising methods such as the one we propose still exists very much.

The objective of this paper is to integrate the strengths of local clustering based methods and those adaptive methods which use neighborhoods of various shapes and sizes. In this way, complicated edge structures and fine image details should be preserved while removing noise from an image. The steering kernel method by \cite{takeda2007kernel} choose small neighborhood near complicated edge structures but gives positive weights to all pixels in the neighborhood. In the proposed method, we use a bigger neighborhood in such regions as well, and preserve the complicated edge structures by employing local clustering in that neighborhood and averaging the intensities only from a meaningful partition of that neighborhood \citep{mukherjee2015image}. The proposed method utilizes explicit edge detection to determine neighborhood shapes and sizes. For pixels close to the edges, we employ local clustering methods to achieve enhanced performance, particularly around complicated edge structures. For pixels far away from the edges, we use a large enough elliptical neighborhood to smooth the region as much as possible. Numerical studies show that the practical performance of the proposed method is superior to both these approaches as well as most state-of-the-art denoising methods.

The organization of the rest of this paper is as follows. Section \ref{method} describes the proposed integrated image denoising method. Useful theoretical results are discussed in Section \ref{theory}. Section \ref{numerical} shows the performance of the proposed method in comparison with a number of state-of-the-art image denoising methods on both the simulated and real images. A few concluding remarks in Section \ref{remarks} finish the paper.

\section{Proposed Methodology}  \label{method}
Assume that the observed image intensities follow the jump regression model
\[z_{ij}=f(x_i,y_i) +\epsilon_{ij} \hspace{0.1in} \mbox{for} \hspace{0.1in} i,j = 1,2,\dots,n, \]
where $\{(x_i,y_j) = (i/n,j/n), \ i,j=1,2,\dots,n\}$ are equally spaced design points or pixels. $\{\epsilon_{i,j} \}$ are independent and identically distributed (i.i.d.) random variables with mean $0$, and variance $\sigma^2$. We assume that $f(x,y)$ is continuous over $ [0,1] \times [0,1] $ except on some curves which we call the {\it jump location curve} (JLC). A point (i,j) is called a singular point if it satisfies at least one of the following conditions:
\begin{itemize}
\item There exists a constant $\zeta_0>0$ such that for any $0 < \zeta < \zeta_0$, the neighborhood of $(i/n,j/n)$ of diameter $\zeta$ is divided into more than two connected regions by the JLCs.
\item There does not exist any constant $\rho_0 >0$ such that there exist two orthogonal lines crossing at $(i/n,j/n)$ and two vertical quadrants formed by these two lines belong to two different regions separated by a JLC in a neighborhood of $(i/n,j/n)$ with diameter $\rho_0$.
\end{itemize}
Detailed explanations are provided by \cite{qiu1997jump}.

\subsection{Detection of edge pixels}
The first step of our proposed method is to detect the edge pixels. There are several methods available in the literature. Any reasonable method can be used in this step. We select the following edge detection method under jump regression framework. Consider the following square neighborhood \[N(x_i,y_i) = \{(x_{i+s},y_{j+t}), \ s,t=-k,-k+1,\dots,0,\dots,k-1,k\}, \] and we fit a least square plane in this neighborhood around $(x_i,y_j)$:
\[
\widehat{z}_{ij}(x,y)= \widehat{\beta}_0^{(i,j)} + \widehat{\beta}_1^{(i,j)}(x-x_i) + \widehat{\beta}_2^{(i,j)}(y-y_j), \hspace{0.1in} (x,y) \in N(x_i,y_j).
\]
We can estimate the gradients of the intensity function using $\left(\widehat{\beta}_1,\widehat{\beta}_2\right)^T$. Considering $B_1$ and $B_2$ to be two pixels nearest to $(i,j)$ along the gradient, whose neighborhoods do not intersect the neighborhood we choose for $(i,j)$. In this paper, we use the following edge detection criterion:
\[
\delta_{ij}=\min\Bigg(||\Vec{v}_{ij}-\Vec{v}_{B_1}||,||\Vec{v}_{ij}-\Vec{v}_{B_2}||\Bigg), \hspace{0.1in}\text{where } \Vec{v}_{ij}=\left(\widehat{\beta}_1^{(i,j)},\widehat{\beta}_2^{(i,j)}\right).
\]
This approach is similar to the one proposed by \cite{qiu1997jump}. $\delta_{ij}$ should be large if and only if $(x_i,y_j)$ is close to a JLC. \cite{qiu1997jump} show that under certain regularity assumptions, $\delta_{ij}>\widehat{\sigma}\sqrt{\frac{\chi^2_{2,\alpha_n}}{kS_x^2}}$, which can be used as an edge pixel detection criterion. Here, $k$ is the window width, $\chi^2_{2,\alpha_n}$ is the $(1-\alpha_n)$ quantile of $\chi^2_2$ distribution, and $S_x^2$ is the sample variance of the $X$-coordinates in the local neighborhood.

\subsection{Local smoothing}
\begin{figure}[ht!]
    \centering
    \includegraphics[width = 5in]{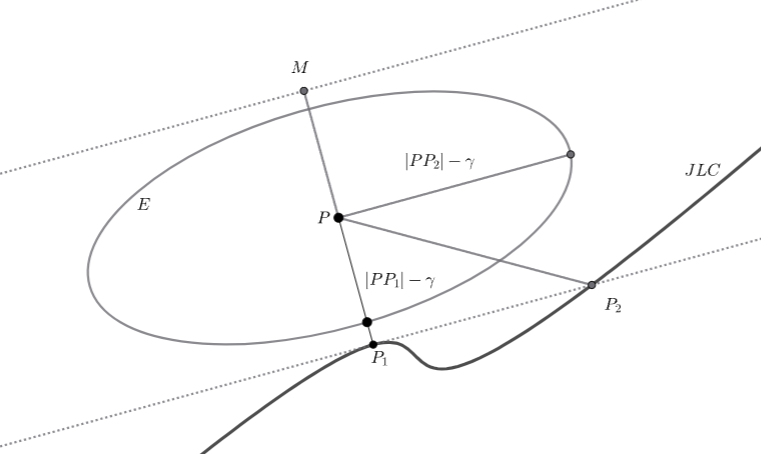}
    \caption{\textbf{Figure demonstrating how elliptical neighborhood around the pixel $P$ is determined.}}
    \label{fig:ellipse}
\end{figure}

In this step of the proposed method, we find an elliptical neighborhood around each pixel of the image such that no detected edge pixels are inside it. To accomplish this, we execute the following steps:
\begin{enumerate}
\item We first find out the nearest edge pixel to a given pixel $P$. Call that edge pixel $P_1$.
\item Next, we consider the point $M$, which is the reflection of the point $P_1$ with respect to the given point. Then, we find the nearest edge pixel from $P$ in the strip perpendicular to the line joining $M$ and $P_1$. Call that edge pixel $P_2$. See Figure \ref{fig:ellipse} for visual explanation.
\item Now, consider the ellipse $E$, centered at $P$, with half-lengths of the major and minor axes given by $|PP_2|-\gamma$ and $|PP_1|-\gamma$, respectively. The tuning parameter $\gamma$ ensures that the selected ellipse neither intersects nor touches any JLC.
\end{enumerate}
The point on the ellipse $E$ along the minor axis that is closest to $P_1$ corresponds to the intersection of the line segment $PP_1$ and $E$. This intersection point, denoted by $P^\ast$, lies exactly $\gamma$ distance away from $P_1$, i.e., $|P^\ast P_1| = \gamma$. If a point $\overline{P}$ on a JLC lies within a distance less than $\gamma$ from the ellipse, then, by the triangle inequality, it follows that $|P\overline{P}| < |PP_2|$, which contradicts the assumption that $P_2$ is the nearest point on the JLC to $P$ within the strip. Therefore, the ellipse maintains a minimum distance of $\gamma$ from all points on any JLC.
See Figure \ref{fig:ellipse} for demonstration.

We then use a suitable kernel function which is non-negative only inside $E$ and zero outside it, and estimate the true intensity of the pixel at $P$ using kernel regression. In kernel regression, the observed pixel intensities are given by:
\vspace{-3mm}
\[z_{\Tilde{p}}=f(\Tilde{p})+\epsilon_{\Tilde{p}},\]
where $f$ is an unknown function and $\tilde{p}$ is the vector of coordinates of a pixel on the image. We assume that the true intensity function $f$ is smooth up to $L$-th order. If $\tilde{p}_i$ is close to $\tilde{p}$, we can use the Taylor series expansion to get:
\begin{align*}
z(\tilde{p}_i) &\approx  z(\tilde{p})+(\nabla z(\tilde{p}))^T (\tilde{p}_i-\tilde{p})+\frac{1}{2} (\tilde{p}_i-\tilde{p})^T\mathcal{H}_{z}(p)(\tilde{p}_i-\tilde{p})+ \dots \\
&=\theta_0+\theta_1^T(\tilde{p}_i-\tilde{p})+\theta_2^Tvech{(\tilde{p}_i-\tilde{p})(\tilde{p}_i-\tilde{p})^T}+...
\end{align*}
where $\theta_0 = z(\tilde{p})$, $\theta_1 = \nabla z(\tilde{p})$, $\theta_2= \frac{1}{2}\Bigg[\frac{\partial^2z(\tilde{p})}{\partial\tilde{p}_1^2},2\frac{\partial^2z(\tilde{p})}{\partial\tilde{p}_1\partial\tilde{p}_2},\frac{\partial^2z(\tilde{p})}{\partial\tilde{p}_2^2}\Bigg]$, $vech
\begin{pmatrix}
a & b\\
b & d
\end{pmatrix} = (a, b ,d)$,
and $\mathcal{H}_{z}(p)$ is the Hessian matrix evaluated at $\tilde{p}$. Therefore, we formulate this as a regression problem and estimate $\theta_0$, $\theta_1$, $\dots$ by minimizing:
\[
\sum_{i=1}^{N_p}\bigg[ z(\tilde{p}_i)-\theta_0-\theta_1^T(\tilde{p}_i-\tilde{p})-\theta_2^Tvech{(\tilde{p}_i-\tilde{p})(\tilde{p}_i-\tilde{p})^T}+...\bigg]^2K\left(\frac{\tilde{p}_i-\tilde{p}}{h}\right),
\]
where $N_p$ is the number of pixels in the local neighborhood, $K$ is a kernel of our choice and $h$ is our chosen bandwidth. We do this only for those pixels where the distance between $P$ and $P_1$ is large enough.

For pixels which are closer to the detected edge pixels, i.e., within $\gamma$ distance of an edge pixel, we are using the clustering method proposed by \cite{mukherjee2015image}. Consider a neighborhood $O(x,y,h_n)$ around one such point $(x,y)$. We can choose a threshold $s$, so that the pixels with intensity value $z_{ij}\leq s$ are in one cluster. Then, we are choosing $s$ such that it maximizes $T(x,y,h_n,s)$, the ratio of between-group and within-group variability of the intensity values.
\[
T(x,y,h_n,s) =  \frac{|O_1(x,y,h_n,s)|(\Bar{z}_1-\Bar{z})^2 + |O_2(x,y,h_n,s)|(\Bar{z}_2-\Bar{z})^2}{\sum\limits_{(x_i,y_j) \in O_1(x,y,h_n,s) } (z_{ij}-\Bar{z}_1)^2 + \sum\limits_{(x_i,y_j) \in O_2(x,y,h_n,s) } (z_{ij}-\Bar{z}_2)^2},
\]
where $O_1(x,y,h_n,s)$ and $O_2(x,y,h_n,s)$ are the two clusters created using threshold $s$, $\Bar{z}_1$ and $\Bar{z}_2$ are the average of intensity values in the respective clusters. We can choose the optimal value of $s$ using: \[
S_0 = \text{arg}\max T(x,y,h_n,s).
\]
After dividing the neighborhood $O(x,y,h_n)$ into two clusters, we estimate the true image intensity at $(x,y)$ using the weighted average of intensity values of the pixels in the cluster containing $(x,y)$. In this paper, we use a similarity measure between pixels as weights, which is given by:
\[
\tilde{W}_{ij} = exp\Bigg( -\frac{||\tilde{O}(x_i,y_j)-\tilde{O}(x,y)||^2_2}{2\hat{\sigma}^2|\tilde{O}(x_i,y_j)|B_n} \Bigg),
\]
where $||\tilde{O}(x_i,y_j)-\tilde{O}(x,y)||^2_2$ is the sum of squares of the pixelwise differences of the intensities between the neighborhoods $\tilde{O}(x_i,y_j)$ and $\tilde{O}(x,y)$, $|\tilde{O}(x_i,y_j)|$ is the number of pixels in the neighborhood $\tilde{O}(x_i,y_j)$, $\hat{\sigma}$ is the estimated standard deviation of errors, and $B_n$ is a tuning parameter for smoothing.

The following flow chart summarizes the proposed integrated image denoising method. Please read the chart along with Figure \ref{fig:ellipse}.

\begin{center}
\begin{tikzpicture}[node distance=1.5cm]

\node (pick) [startstop] {\small Pick a pixel $P$ from the image.};
\node (edge) [startstop,right of= pick,xshift=3cm]{\small Find the nearest edge pixel $P_1$, and calculate $|P P_1|$.};
\draw [arrow] (pick) -- (edge);
\node (dist) [startstop,right of=edge,xshift=3cm]{\small $|P P_1|<\gamma$};
\draw [arrow] (edge)--(dist);

\node (smoothcase) [startstop,below of=dist,yshift=-2cm]{\small Find nearest pixel  $P_2$ in the strip perpendicular to $P_1M$, where $M$ is reflection of $P_1$ with respect to $P$.};
\draw [arrow] (dist) -- node[anchor=east] {no} (smoothcase);
\node (edgecase) [startstop,right of=smoothcase,xshift=3cm]{\small Smooth using clustering method.};
\draw [arrow] (dist) -| node[anchor=west] {yes} (edgecase);
\node (ellipse) [startstop,left of=smoothcase,xshift=-3cm]{\small Construct an ellipse with $P$ as center, and $|PP_1|-\gamma$, $|PP_2|-\gamma$ as half-lengths of the axes.};
\draw [arrow] (smoothcase)--(ellipse);
\node (ellipse2) [startstop,left of=ellipse,xshift=-3cm]{\small Smooth in the neighborhood E around P using kernel regression.};
\draw [arrow] (ellipse) -- (ellipse2);

\end{tikzpicture}
\end{center}


\subsection{Selection of procedure parameters}

For our proposed algorithm, we need four parameters: the bandwidth and cutoff for the edge detection step, and bandwidth parameters for the local clustering step and smoothing step, and one tuning parameter for the smoothing step after local clustering. In our implementation, we introduce only one new parameter, the maximum possible axis length of the elliptical neighborhood. From numerical experience, we set this parameter to $6$ pixels for images having resolution less than $100 \times 100$, and $10$ pixels for images with higher resolution. Similarly, the bandwidth for the clustering method is chosen to be the  same as the minimum distance $\gamma$ to apply the method, which we choose to be $3$ pixels for images with resolution less than $100 \times 100$, and $5$ pixels otherwise. The rest of the parameters are chosen according to the guidelines mentioned by \cite{mukherjee2015image} and \cite{qiu1997jump}.

\subsection{Computational efficiency}
As our algorithm integrates two different algorithms with distinct computation times, comparison of their computational efficiencies is essential. Suppose, $\mathcal{C}_1$ denotes the computation time of the edge detection algorithm, $\mathcal{C}_2$ and $\mathcal{C}_3$ denote the computation times of the regression method and the clustering method per pixel, respectively. Hence, the total computation time $\mathcal{C}_{new}$ of the integrated algorithm is given by: $$\mathcal{C}_{new}=\mathcal{C}_1+\sum\limits_{x,y}(P(x,y)\mathcal{C}_2 + (1-P(x,y))\mathcal{C}_3),$$
where $P(x,y)$ represents the probability that the nearest edge pixel to the point $(x,y)$ lies at a distance greater than $\gamma$. Noting that $\mathcal{C}_1+n^2\min(\mathcal{C}_2,\mathcal{C}_3)\leq\mathcal{C}_{new} \leq\mathcal{C}_1+n^2\max(\mathcal{C}_2,\mathcal{C}_3)$,
we conclude that the computation time of the proposed integrated method lies between those of a single iteration of the two baseline algorithms. Moreover, the regression based smoothing by \cite{takeda2007kernel} involves multiple iterations, hence its computation time is $\mathcal{T}n^2 \mathcal{C}_2$, $\mathcal{T}$ being the number of iterations. The computation time of the local clustering method by \cite{mukherjee2015image} is $\left(\mathcal{C}_1 + n^2 \mathcal{C}_3\right)$ with $\mathcal{C}_3$ being slightly larger than $\mathcal{C}_2$, but of similar order with respect to $n$. Hence, for sufficiently large $\mathcal{T}$ and $n$, $\mathcal{C}_{new}$ is smaller than both $\mathcal{T}n^2 \mathcal{C}_2$ and $\left(\mathcal{C}_1 + n^2 \mathcal{C}_3\right)$.

Although the proposed integrated algorithm involves a two-step procedure, apart from initial edge detection, the rest of the algorithm can be performed independently on each pixel. Therefore, the second step of this integrated algorithm can be parallelized. Furthermore, by choosing a computationally lightweight edge detection algorithm, we can significantly reduce the overall computation time.

\section{Statistical Properties} \label{theory}
In this section, we discuss important theoretical properties of the proposed integrated procedure.
The following result justifies its asymptotic consistency.

\textbf{Theorem 1:}
Assume that f has continuous first-order derivatives over $(0,1) \times (0,1)$ except on JLCs, the first order one-sided derivatives exist at the JLCs, $\epsilon_{ij}$ are independent and have distribution $N(0,\sigma^2)$ with $0<\sigma^2<\infty$. If we have ${h}^\ast_n= O(n^\beta)$ with $\beta <1$, $\alpha_n = O(1/\log \log n)$, $h_n = o(1)$, $1/nh_n = o(1)$,  and the clustering bandwidth $\gamma_n<h_n\text{ with } \gamma_n=O(n^{-1}\log n)$, where $h^\ast_n$ is the bandwidth parameter for edge detection, and $h_n$ is the bandwidth parameter for smoothing. Then, for any non-singular point $(x,y)$ with the property $d((x,y),(x^*,y^*))>\epsilon$ for all singular points $(x^*,y^*)$, we have $\widehat{f}(x,y)=f(x,y)+ O(\max(h_n,\gamma_n^{1/2}))$. where $\widehat{f}(x,y)$ is the estimate of $f(x,y)$ produced by the proposed integrated method.

Sketches of the proof of the above theorem are based on \cite{mukherjee2015image} and \cite{qiu1997jump}. We provide the details in the Appendix.

{\bf Remark:} Although we make normality assumption of the additive noise for simplicity of the proof, it is not critically important. As long as the density of the error distribution is symmetric around zero, and has finite moments at least up to a certain order, point-wise convergence of $\widehat{f}(x,y)$ holds. If these assumptions on the error distribution do not hold, then we need to use other techniques, such as noise-adaptive transforms or pre-processing for non-Gaussian noise.



\section{Numerical Studies} \label{numerical}
In this section, we present several numerical experiments where we study the performance of the proposed integrated method which we call NEW, in comparison with four competing methods. We consider three noise levels representing low, medium, and high, and various state-of-the-art denoising techniques for performance comparison. We use integrated root mean square error (IRMSE) to measure the performance of various image denoising methods which is defined as:
\begin{equation*}
    \text{IRMSE}= E\bigg(\sqrt{\int\limits_{z\in \Omega}(\widehat{f}(z)-f(z))^2}\bigg), \text{ where } \Omega=[0,1]\times[0,1].
\end{equation*}
In our numerical studies, we estimate IRMSE by performing multiple instances of repeated independent simulations, and calculating the sample mean of root mean square error (RMSE) values:
 \begin{equation*}
     \widehat{\text{IRMSE}}= \frac{1}{L}\sum\limits_{\ell=1}^L\bigg(\sqrt{\sum\limits_{(x_i, y_j)}(\widehat{f}_{\ell}(x_i, y_j)-f(x_i, y_j))^2}\bigg).
 \end{equation*}

\subsection{A brief description of the competing methods}
Below we provide a brief overview of the competing methods that we select for the numerical study.

\textbf{(i) Method based on local pixel clustering:} The clustering method proposed by \cite{mukherjee2015image} is the one we use in our integrated algorithm when the given pixel is close to a jump location curve in the given image. We include it in our comparison study to demonstrate that our proposed method is at least as good as this method. Moreover, the proposed method can denoise a given image in a much lesser time by eliminating the need to perform clustering in smoother regions. We choose the window width based on empirical optimality while selecting other parameters as suggested by \cite{mukherjee2015image}. For implementation, we use the computer codes provided in the supplementary materials of \cite{mukherjee2015image}.

\textbf{(ii) Steering kernel algorithm:} The steering kernel method proposed by \cite{takeda2007kernel} uses an adaptive approach to denoise a given image. Instead of calculating the edge pixels directly, it uses gradient information to find a direction which should be parallel to the nearby edge, and determines an elliptical neighborhood elongated in that direction. Since we incorporate a similar approach near the jump location curves of the given image, it is natural to include this method in our comparison study. It is to be noted that the steering kernel method is an iterative method, i.e., it applies the denoising method several times until the estimated mean squared error is small enough, whereas our proposed method is not iterative, and hence computationally less expensive. We select the parameters of the steering kernel method as suggested by \cite{takeda2007kernel}. We use the computer codes provided by \cite{Steering} for implementation.

\textbf{(iii) Jump preserving local linear kernel (JPLLK) method:} This procedure proposed by \cite{jpllk2009} involves a two-step process for surface reconstruction using local linear kernel smoothing. At a given point, a plane is fitted using local linear kernel smoothing to find a standard local linear kernel estimator of the image intensity function. The surrounding area is split into two sections by a line that passes through the point and is perpendicular to the gradient of the fitted plane. In each section, a half-plane is fitted using local linear kernel smoothing, resulting in two one-sided estimators of the image intensity function at that point. Subsequently, the best estimate among the two one-sided estimates, and the standard two-sided estimate is chosen as the final estimate at that point. This process is iterated twice to further smooth the surface. We select the bandwidth parameter of this smoothing procedure to get numerically optimal performance. \cite{DRIP} provide computer codes for execution.

\textbf{(iv) Method by total variation regularization (TV):} This is a very popular method used extensively in various applications. In this method, the total variation of an image is minimized using certain constraints related to the statistics of the noise. \cite{RUDIN1992259} show that reducing the total variation of an image removes noise while preserving essential structural details. Moreover, \cite{chambolle2011first} develop a first order primal dual algorithm based on this method that works well for image denoising purposes alongside various other applications. We select parameters following the suggestion by \cite{chambolle2011first}. We use the computer codes provided by \cite{tvR}.

It is to be noted that we do not consider any deep learning based denoising algorithms (e.g., \cite{deepdenoising}) as the competitor of the proposed method because they require a large set of images for training or learning. Hence, such type of denoising methods cannot be easily integrated with image monitoring procedures which itself is a computationally heavy problem. Moreover, the problem setup that we consider does not involve any training or learning on a large class of images. Although the self-supervised learning methods \citep{Self-to-self2020, Self-supervised2024} simulate training data from the given noisy image, computational complexity still remains as an issue if we wish to use the denoising method in subsequent analysis such as image monitoring. Therefore, comparing the proposed method with deep learning based denoising algorithms is rather unfair.

\subsection{Simulation studies}\label{sim}
\begin{figure}[htb!]
    \centering
    \includegraphics[width=6.5in]{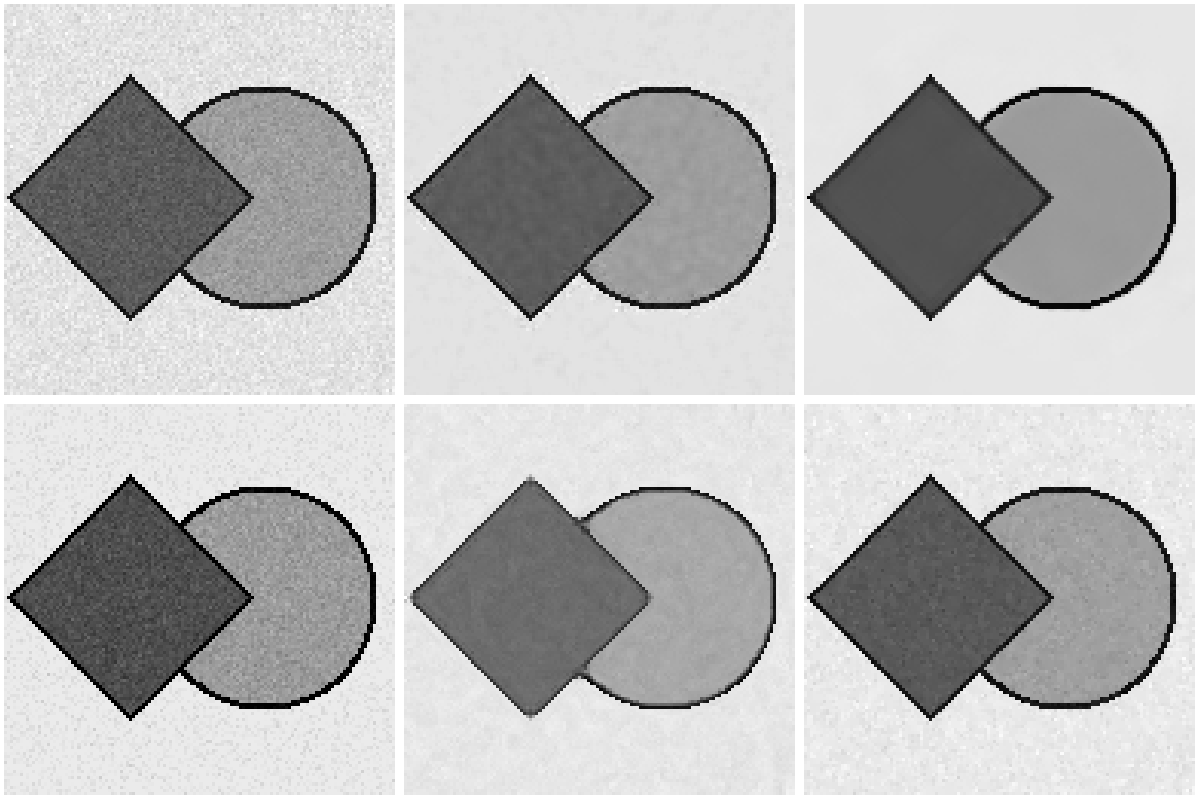}
    \caption{Denoising performance on the simulated image with additive Gaussian noise of $20$ standard deviation. Panels from the top-left to the bottom-right show the noisy image, and the denoised images by the proposed method, clustering algorithm, steering kernel, JPLLK, and TV, respectively.}
    \label{fig:sqcirc_performance}
\end{figure}

We first construct a simple simulated image consisting of a square and a part of a circle. This image has various types of edges and edge structures such as linear edge, corner, intersection of a line with a curve and so forth. A noisy version of the simulated image is presented in the $(1, 1)$-th panel of Figure \ref{fig:sqcirc_performance}. We choose such an image to demonstrate the performance of the proposed integrated method in comparison with its competitors. The outcomes are presented in Table \ref{tab:sim_mse_performance} and Figure \ref{fig:sqcirc_performance}.

From Figure \ref{fig:sqcirc_performance}, we see that our method visually performs better than its competitors. Although the difference between our algorithm and the clustering algorithm is minimal, we can see that the clustering algorithm over-smooths the image, i.e., smudges the image around edges while our method performs better in that regard. This is due to the fact that clustering step in our integrated method selects a smaller neighborhood.

\begin{table}[htb]
\footnotesize
\centering
        \begin{tabular}{ |c|c|c|cccc| }
        \hline
        \multicolumn{2}{|c|}{}& \multicolumn{1}{c|}{} & \multicolumn{4}{c|}{Competing Methods} \\
        \hline
         Resolution & Noise & Proposed & Steering Kernel  & Clustering &JPLLK &TV  \\
        \hline
            &  20 & 11.79 (0.317) &  15.97 (0.300)   & 12.81 (0.991)& 16.09 (0.278) & 18.87 (0.252)\\
        {$64 \times 64$}& 10  & 5.47 (0.141)  & 7.98 (0.150)  & 8.28 (0.138) & 19.34 (0.535) & 8.93 (0.134) \\
        & 5  &  3.36 (0.081) &  3.98 (0.075) & 7.24 (0.057) & 28.19 (0.857) & 4.35 (0.104) \\
        \hline
            &  20 & 9.22 (0.119) &  16.53 (0.123)   & 9.32 (0.105) & 16.01 (0.112) & 18.75 (0.132)\\
        {$128 \times 128$}& 10  & 4.43 (0.039)  & 8.26 (0.061)  & 7.96 (0.061) & 12.10 (0.062) & 8.82 (0.081)  \\
        & 5  &  2.30 (0.035) & 4.11 (0.031)  & 7.42 (0.031) & 10.97 (0.035) & 4.24 (0.099) \\
        \hline
        \end{tabular}    
    \caption{Performance comparisons for simulated images using estimated IRMSE.}
    \label{tab:sim_mse_performance}
\end{table}

Table \ref{tab:sim_mse_performance} shows the performance of the concerned methods in terms of IRMSE. We run each of the concerned methods on images with different instances of noise $30$ times independently, and calculate the sample mean of RMSE as our estimated IRMSE. The numbers in parentheses show the standard error of the estimates. Table \ref{tab:sim_mse_performance} reflects the same phenomenon we observe from Figure \ref{fig:sqcirc_performance}. The numerical performance of the proposed method is quite similar to the clustering method, but much better than other competitors. It is expected to be similar to the clustering method as the proposed method is an improvement in terms of computational complexity. As seen from Table \ref{tab:sim_mse_performance} and Figure \ref{fig:sqcirc_performance}, the proposed integrated method performs slightly better in the regions where the true image is smooth. This results in small improvement in the estimated IRMSE. In comparison with the steering kernel algorithm, the proposed method is better both visually and with respect to the estimated IRMSE values. JPLLK performs quite well in terms of smoothing visually, but it performs poorly where two jump location curves intersect which is reflected in the estimated IRMSE values. While the TV method performs well in low noise levels, it performs poorly in high noise levels. Our proposed integrated method performs better in comparison to its competitors especially when the noise level is high.

The performance of the proposed integrated method along with the baseline methods are very similar when the noise distribution is non-Gaussian but symmetric around zero with finite first few moments. Performances of the jump regression based image denoising methods have been explored in the literature in presence of such kinds of non-Gaussian noise, and these performances are very similar to the Gaussian noise cases \citep{mukherjee2015image, QiuMukherjee2010}. These observations are in line with our remark after Theorem 1.

\subsection{Performance comparison on real images}\label{real}
\begin{figure}[htb!]
    \centering
    \includegraphics[width=6.5in]{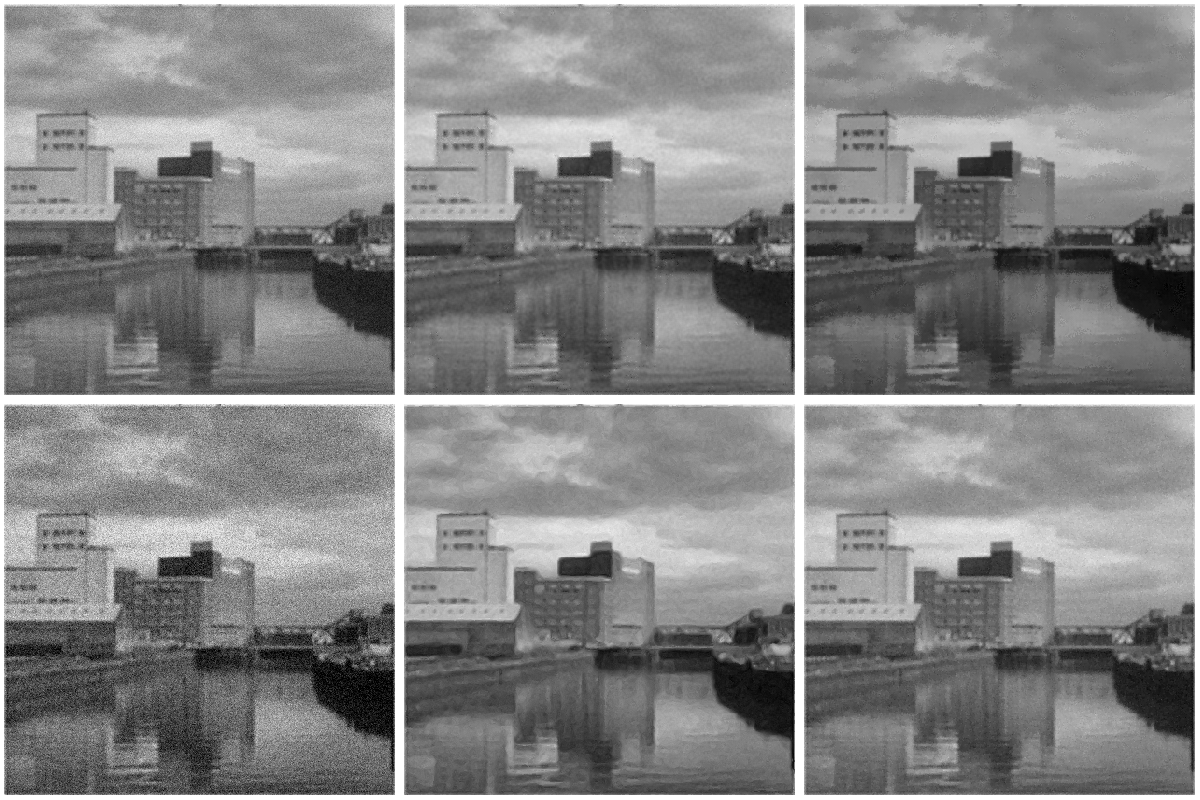}
    \caption{Performance comparison on the cityscape image with additive Gaussian noise of 10 standard deviation. Panels from the top-left to the bottom-right show the noisy image, and the denoised images by the proposed method, clustering algorithm, steering kernel, JPLLK, and TV, respectively.}
    \label{fig:cityplot}
\end{figure}

\begin{figure}[htb!]
    \centering
    \includegraphics[width=6.5in]{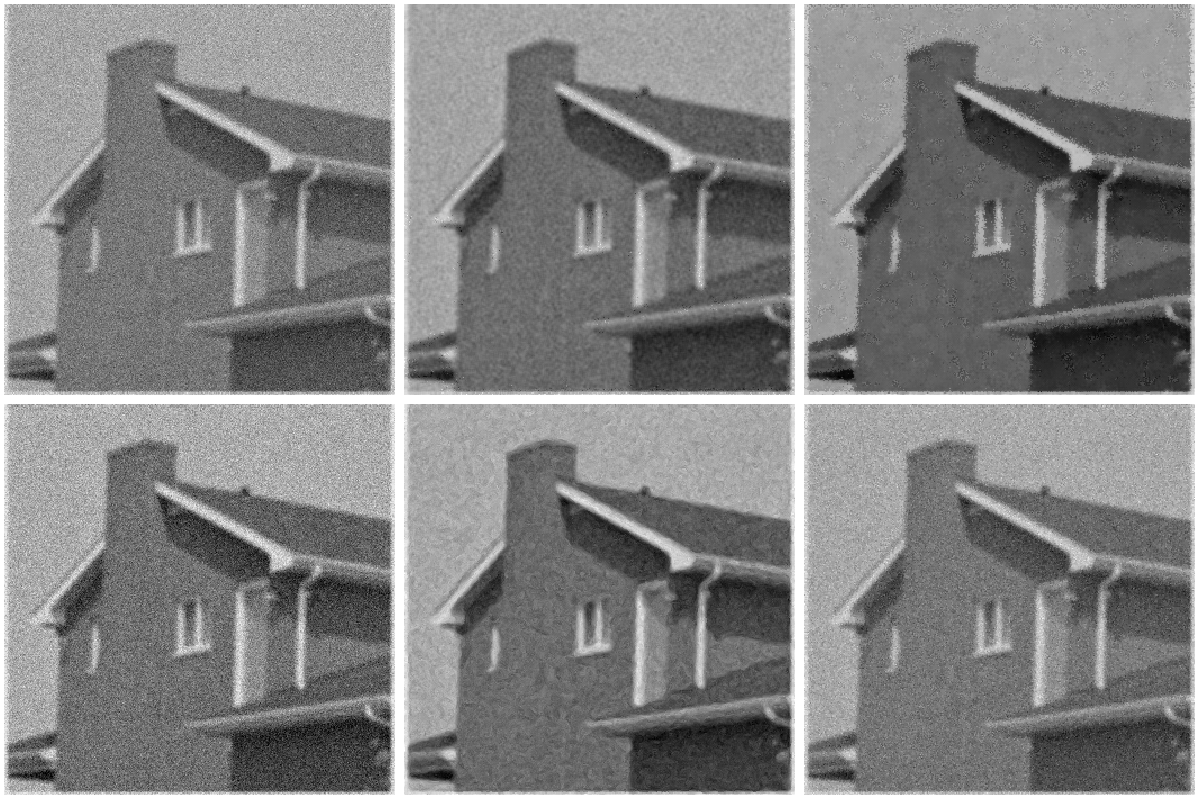}
    \caption{Performance comparison on the house image with additive Gaussian added noise of 10 standard deviation. Panels from the top-left to the bottom-right show the noisy image, and the denoised images by the proposed method, clustering algorithm, steering kernel, JPLLK, and TV, respectively.}
    \label{fig:house}
\end{figure}

In this subsection, we consider two real images, one of a cityscape, and another of a house to evaluate the performance of the denoising methods. Noisy versions of the images are presented in the first panel of Figures \ref{fig:cityplot} and \ref{fig:house}. The outcomes are presented in Table \ref{tab:real_mse_performance}, and Figures \ref{fig:cityplot} and \ref{fig:house}.

\begin{table}[htb!]
\footnotesize
    \begin{center}
        \begin{tabular}{ |c|c|c|cccc| }
        \hline
        \multicolumn{2}{|c|}{}& \multicolumn{1}{c|}{} & \multicolumn{4}{c|}{Competing Methods} \\
        \hline
         Image & Noise & Proposed & Steering Kernel  & Clustering &JPLLK&TV  \\
        \hline
            &  20 & 7.92 (0.029) &  15.34 (0.069)   & 8.52 (0.059)& 9.12 (0.042) & 18.62 (0.033)\\
        City by water& 10  & 4.69 (0.016)  &  9.96 (0.017) & 4.87 (0.026)& 7.81 (0.025) & 8.63(0.026) \\
        & 5  & 2.82 (0.010)  & 5.33 (0.012)  & 2.88 (0.011)& 7.42 (0.015)&3.84 (0.012)  \\ \hline
        &  20 & 7.65 (0.028) &  15.22 (0.073)   & 8.63 (0.065)& 8.35 (0.046) & 18.66 (0.044)\\
        House & 10  &  4.37 (0.017) &  9.78 (0.020) &  4.60 (0.031)& 6.86 (0.024) & 8.86 (0.029)\\
        & 5  & 2.51 (0.009)  & 4.99 (0.139)  &  2.49 (0.018)& 6.40 (0.012) & 3.87 (0.02)\\
        \hline   
        \end{tabular}    
    \end{center}
    \caption{ Performance comparisons for real images using estimated IRMSE.}
    \label{tab:real_mse_performance}
\end{table}

\vspace{-3mm}

\vspace{2mm}

From Table \ref{tab:real_mse_performance}, we observe similar patterns as in the case of the simulated image. The proposed algorithm outperforms the steering kernel method, JPLLK, and TV in terms of estimated IRMSE for both of the images across all three noise levels that we consider. The performance of the proposed algorithm and the local clustering method continue to be similar both visually and in terms of estimated IRMSE. In fact, the estimated IRMSE of the proposed method is smaller than that of the local clustering method in several cases. The steering kernel algorithm and TV method perform reasonably well when the noise level is low. The proposed method outperforms them in all cases with bigger margins at higher noise levels. In contrast, the performance of JPPLK is good at higher noise level but it perform rather poorly at lower noise level. Therefore, the proposed integrated algorithm performs well in nearly all situations without compromising the smoothing performance in the continuity regions while being less computationally expensive than the local clustering method.

Another important aspect to note is that the proposed algorithm performs better when the underlying edge structures of images are defined more clearly. This is because it relies heavily upon the edge detection algorithm we use. Therefore, in cases where edge pixels are not detected well, the proposed method ends up smoothing those edges. However, when the noise level is very high, this issue is usually overshadowed due to heavy noise in comparison with the jump size.

\section{Concluding Remarks} \label{remarks}
In this paper, we propose an integrated image denoising procedure which detects the edge pixels first, and then proceed to denoise using either adaptive kernel regression or local clustering depending on local information regarding edges. This paper shows better performance when compared to similar methods already in the literature. We also provide a theoretical justification that the image intensity function of the denoised image converges point-wise to the true image intensity function for large image size, under mild regularity assumptions. However, there are scopes for extending the proposed approach. In this paper, we consider the local clustering algorithm developed by \cite{mukherjee2015image} as baseline for denoising around edges. This can be improved by using more recent image denoising methods. Though it is restricted by the framework of the problem that we consider in this paper, one may expand this approach to use other types of denoising methods such as wavelet based methods, diffusion based methods, and many more. Another direction of potential improvement is the number of methods we can stitch together using this approach. In this paper, we divide the problem into just two parts, namely smoothing near the edges, and smoothing far away from the edges. However, in future research, one may divide the problem into more parts, and thus have an assortment of methods that perform well in specific scenarios which altogether produce superior overall result. One immediate application of the proposed image denoising method is in designing techniques for image comparison and image monitoring.




\bibliography{bibliography}

\appendix

\section{Appendix} \label{appen}
Let us define the following notations:
$$
\begin{aligned}
\Omega_{\epsilon} &=[\epsilon, 1-\epsilon] \times[\epsilon, 1-\epsilon], \\
S_{\epsilon} &=\left\{(x, y):(x, y) \in \Omega, d_{E}\left((x, y),\left(x^{*}, y^{*}\right)\right) \leq \epsilon \text { for a singular point }\left(x^{*}, y^{*}\right) \in D\right\}, \\
\Omega_{\bar{S}, \epsilon} &=\Omega_{\epsilon} \backslash S_{\epsilon}.
\end{aligned}
$$
Next, we proceed with the following Lemmas:

{\bf Lemma 1:} Assume that $f(x, y)$ has continuous first order partial derivatives over $(0,1) \times(0,1)$ except on the JLCs, and its first order one-sided partial derivatives at the JLCs exist. Assume that $\lim _{n \rightarrow \infty} h^\ast=\infty$ and $\lim _{n \rightarrow \infty} h^\ast / n=0$, where $h^\ast$ is the window width. If there is no jump in $N\left(x_{i}, y_{j}\right)$, then
$$
\hat{\beta}_{1}^{(i, j)}=f_{x}^{\prime}\left(x_{i}, y_{j}\right)+O\left(\frac{n \sqrt{\log \log h^\ast}}{{h^\ast}^{2}}\right) \quad \text{a.s. }
$$
and
$$
\hat{\beta}_{2}^{(i, j)}=f_{y}^{\prime}\left(x_{i}, y_{j}\right)+O\left(\frac{n \sqrt{\log \log k}}{k^{2}}\right) \quad \text{a.s. }
$$

If $\left(x_{i}, y_{j}\right)$ is on a JLC and it is not a singular point, then

$$
\hat{\beta}_{1}^{(i, j)}=f_{x}^{\prime}\left(\tilde{x}_{i}, \tilde{y}_{j}\right)+h_{1}^{(i, j)} C(i, j)+\Gamma_{1} C_{x}(i, j)+O\left(\frac{n \sqrt{\log \log h^\ast}}{{h^\ast}^{2}}\right) \quad \text {a.s. }
$$
and
$$
\hat{\beta}_{2}^{(i, j)}=f_{y}^{\prime}\left(\tilde{x}_{i}, \tilde{y}_{j}\right)+h_{2}^{(i, j)} C(i, j)+\Gamma_{2} C_{y}(i, j)+O\left(\frac{n \sqrt{\log \log h^\ast}}{{h^\ast}^{2}}\right) \quad \text {a.s.,}
$$
where $\left(\tilde{x}_{i}, \tilde{y}_{j}\right)$ is some point around $\left(x_{i}, y_{j}\right)$ on the same side of the JLC as $\left(x_{i}, y_{j}\right)$, and the distance between $\left(\tilde{x}_{i}, \tilde{y}_{j}\right)$ and $\left(x_{i}, y_{j}\right)$ converges to zero; $C(i, j), C_{x}(i, j)$, and $C_{y}(i, j)$ are the jump magnitudes of $f(x, y)$ and its first order $x$ and y partial derivatives; $h_{1}^{(i, j)}$ and $h_{2}^{(i, j)}$ are two constants satisfying
$$
\sqrt{\left(h_{1}^{(i, j)}\right)^{2}+\left(h_{2}^{(i, j)}\right)^{2}}=O(n / h^\ast),
$$
$\Gamma_{1}$ and $\Gamma_{2}$ are two constants between $-1$ and $+1$.

{\bf Lemma 2:} Besides the conditions stated above, if there are only finite number of singular points on JLCs, and $\alpha_{n}$ is chosen such that 
\begin{itemize}
    \item $\lim_{n \rightarrow \infty} \alpha_{n}=$ $0 ;$
    \item $\lim_{n \rightarrow \infty} \log \left(\alpha_{n}\right) / \log (\log (h^\ast))=-\infty$;
    \item and $\lim_{n \rightarrow \infty} \log \left(\alpha_{n}\right) / {h^\ast}^{2}=0$,
\end{itemize}
the point-set of the detected jumps is strongly consistent in the Hausdorff distance and the convergence rate is $O\left(n^{-1} \log (n)\right)$. \\

{\bf Lemma 3:} Assume that $f$ has continuous first-order derivatives over $(0,1) \times(0,1)$ except on the JLCs, its first order derivatives have one-sided limits at non-singular points of the JLCs, $\left\{\varepsilon_{i j}\right\}$ are i.i.d. and have the common distribution $N\left(0, \sigma^{2}\right)$, where $0<\sigma<\infty$, $h_{n}=o(1)$, $1 / n h_{n}=o(1)$, $u_{n}=\kappa+\delta$, where $\kappa=\left(\frac{\phi^{2}(0)}{\Phi(0)(1-\Phi(0))}\right) /\left[1-\left(\frac{\phi^{2}(0)}{\Phi(0)(1-\Phi(0))}\right)\right]$, $\delta$ is any positive number; $\phi$ and $\Phi$ are the probability density function and the cumulative distribution function of the $N(0,1)$ distribution, respectively. If we further assume that $B_{n}=$ $O\left(h_{n}^{1 / 2}\right)$, then for any non-singular $(x, y) \in \Omega_{\bar{S}, \epsilon}$, we have $\widehat{f}(x, y)=f(x, y)+O\left(h_{n}^{1 / 2}\right)$, a.s.

Proofs of Lemmas 1 and 2 are provided in \cite{qiu1997jump} whereas \cite{mukherjee2015image} provide a sketch of the proof of Lemma 3.

From Lemmas 1 and 2, we can say that if we choose $\alpha_n=O(1/log(log(n))$ and $ h^\ast = O(n^\beta)$ where $\beta <1$, all of the above conditions are satisfied for edge detection.

Suppose that we select the bandwidth $h^\ast_n$ and the significance level $\alpha_n$ for edge detection; the bandwidths $h_n$ and $\gamma_n$ for smoothing and clustering, respectively, and the tuning parameter $B_n$ for smoothing, according to Lemma 3. Now that all required conditions for the parameters of above lemmas are satisfied, we now proceed to prove Theorem 1.

\vspace{5mm}

\noindent \textbf{Proof of Theorem 1:} 
\begin{proof}
Let $J$ denote the points on JLCs and $\hat{J_n}$ denote the detected edge pixels.
Suppose, we consider the point $(\tilde{x},\tilde{y})$.

Suppose that the point is not on $J$. Then, as $\mathcal{H}(J,\hat{J}_n)\rightarrow 0$, for large $n$, it is not in $\hat{J}_n$ with probability 1. Therefore, without any loss of generality, we can assume that it is not on $\hat{J}_n$. Suppose that its nearest point on $\hat{J}_n$ is $(x^\ast,y^\ast)$.

There exists $n^\ast$ such that for all $n>n^\ast$, we have $\mathcal{H}(J,\hat{J_n})<\frac{\gamma_n}{2}$.
Hence, we can conclude that $\forall n> n^\ast$, there is a circle around $(\tilde{x}, \tilde{y})$ of radius $d((x^\ast,y^\ast),(\tilde{x},\tilde{y}))-\gamma_n$ which does not intersect or touch $\hat{J_n}$.

Now, for choosing the second closest point in our algorithm (point $P_2$ in Figure \ref{fig:ellipse}), we are doing the same as above while restricting the image in that strip. Therefore, we consider $\hat{\Omega}=\Omega\cap S$ and $\hat{J}^{'}_n=\hat{J}_n\cap S$, where $S$ is the strip chosen using the first step. Due to the fact that the ellipse we define is completely inside the strip and using the same argument as choosing the first point we can say that the ellipse we are choosing has no point which are on any JLC. Hence, $f(x,y)$ has continuous first order derivatives inside the elliptical neighborhood $E$. As the maximum possible axis half-length, $h_n\rightarrow 0$, the Taylor series converges. Therefore, $\hat{f}(\tilde{x},\tilde{y})\rightarrow f(\tilde{x},\tilde{y}) $ i.e., $\widehat{f}(\tilde{x}, \tilde{y})=f(\tilde{x}, \tilde{y})+O\left(h_{n}-\gamma_n\right)=f(\tilde{x}, \tilde{y})+O\left(h_{n}\right)$, a.s.

On the other hand, suppose that, $(\tilde{x},\tilde{y})$ is on J. Then, as $\mathcal{H}(J,\hat{J}_n)\rightarrow 0$, we can argue that $d((\tilde{x},\tilde{y}),\hat{J}_n)<\gamma_n$ for large enough $n$. Therefore, for large enough $n$, we execute the method proposed by \cite{mukherjee2015image}. By Lemma 3, we get $\widehat{f}(\tilde{x}, \tilde{y})=f(\tilde{x}, \tilde{y})+O\left(\gamma_n^{1 / 2}\right)$ a.s.

Theorem 1 is thus proved. \hfill
\end{proof}

\end{document}